\documentstyle[12pt,graphicx,caption2]{article}

\begin{document}

\title{On the multistep processes in the nuclei}
\author{V.I.Nazaruk\\
Institute for Nuclear Research of the Academy of Sciences of Russia,\\
60th October Anniversary Prospect 7a, 117312 Moscow, Russia.*}

\date{}
\maketitle
\bigskip

\begin{abstract}
The intermediate-state interaction and structure of amplitudes of complicated
processes in medium (decays, reactions and the $n\bar{n}$ transitions) are
studied. It is proposed to use the branching ratio of channels of free-space
hadron-nucleon interaction as a test in the construction and verification of
the models. The corresponding formulas for the processes in the medium are
obtained. The connection between particle self-energy and amplitudes of
subprocesses are analyzed as well.
\end{abstract}

\vspace{5mm}

\vspace{5mm}

\vspace{5cm}
*E-mail: nazaruk@inr.ru

\newpage

\setcounter{equation}{0}
In this paper the structure of amplitudes of complicated processes in nuclei is
studied by the example of the decay in the medium. We briefly summarize the
well-known results relating to the unitarity and analyze the value and role of
intermediate particle self-energy. This point is important since an additional
self-energy, as a rule, leads to suppress the process. The expressions for the
branching ratio of decay channels in the low-density approximation are derived
as well. The formulas obtained have a clear physical meaning (we emphasize this
fact), which enables one to verify and correct the model. It is shown that the
models of realistic processes in the nuclei should reproduce the branching ratio of
channels of the corresponding free-space hadron-nucleon interactions. This can be
considered as necessary condition for the correct model construction.

Let us consider a free-space decay $a\rightarrow \pi ^0\bar{n}$, for example,
$\bar{\Lambda }\rightarrow \pi ^0\bar{n}$. For a decay in the medium the annihilation
\begin{equation}
a\rightarrow \pi ^0+\bar{n}\rightarrow \pi ^0+M
\end{equation}
and scattering
\begin{equation}
a\rightarrow \pi ^0+\bar{n}\rightarrow \pi ^0+\bar{n}
\end{equation}
channels take place. Here $M$ are the annihilation mesons, $\bar{n}\rightarrow M$
and $\bar{n}\rightarrow \bar{n}$ imply the annihilation and scattering of $\bar{n}$
in the medium, respectively. In the following the antineutron and $a$-particle are
assumed non-relativistic and spinless.

Let $\Gamma _a$, $\Gamma _s$ and $\Gamma _t$ be the widths of the channels (1),
(2) and the total width of the decay $a\rightarrow \pi ^0\bar{n}$ in the medium,
respectively. For a decay in nuclear matter $\Gamma _s\approx 0$ because the
$\bar{n}$ annihilates in a time $\tau _a\sim 1/\Gamma $, where $\Gamma $ is the
annihilation width of $\bar{n}$ in the medium. The $\Gamma _s$ is taken into
account since the $\rho $-dependence of the results is considered.

The interaction Hamiltonian is
\begin{equation}
{\cal H}_I={\cal H}_d+{\cal H},
\end{equation}
\begin{equation}
{\cal H}_d=g\Psi ^+_{\bar{n}}\Phi \Psi _a+{\rm H.c.},
\end{equation}
where ${\cal H}_d$ is the Hamiltonian of the decay $a\rightarrow \pi ^0\bar{n}$,
${\cal H}$ is the Hermitian Hamiltonian of the $\bar{n}$-medium interaction
taken in the general form. We focus on the intermediate-state interaction of the
$\bar{n}$ and so the $\pi^0$-medium interaction is inessential for us.

The total decay width $\Gamma _t$ can be obtained from the unitarity condition:
\begin{equation}
\Gamma _t=\frac{1}{T_0}(1-\mid S_{ii}\mid ^2)\approx \frac{1}{T_0}2ImT_{ii},
\end{equation}
$S=1+iT$. Here $T_0$ is the normalization time, $T_0\rightarrow \infty $.
The on-diagonal matrix element $T_{ii}$ is shown in Fig. 1a. It involves the
full in medium antineutron propagator $G_m$ which should be calculated through
the Hermitian Hamiltonian ${\cal H}$: $G_m=G_m({\cal H})$. Since
$G_m({\cal H})$ contains the annihilation loops, the realization of this scheme
is extremely complicated.

\begin{figure}[h]
  {\includegraphics[height=.25\textheight]{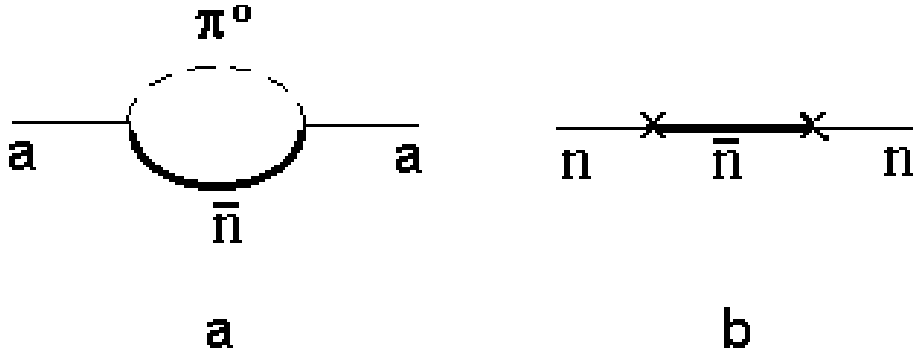}}
  \caption{{\bf a} The on-diagonal matrix element $T_{ii}$ corresponding to the
decay $a\rightarrow \pi ^0\bar{n}$ in the medium. {\bf b} Same as {\bf a} for the
$n\bar{n}$ transition.}
\end{figure}

In the optical model
\begin{equation}
{\cal H}\rightarrow  ({\rm Re}U_{\bar{n}}+i{\rm Im}U_{\bar{n}})
\Psi ^+_{\bar{n}}\Psi_{\bar{n}},
\end{equation}
where $U_{\bar{n}}$ is the optical potential of the $\bar{n}$. Then $G_m({\cal H})
\rightarrow G_m(U_{\bar{n}})$. If $U_{\bar{n}}={\rm const}$,
\begin{equation}
G_m(U_{\bar{n}})=-\frac{1}{\hat{p}_{\bar{n}}-m-U_{\bar{n}}}
\end{equation}
($m$ is the antineutron mass) which greatly simplifies the calculation.

However, in [1,2] it was shown that model (6) cannot be used for the calculation
of $\Gamma _t$ by means of (5). This is because Eq. (5) follows from the unitarity
condition and can be used only for the unitary $S$-matrix, whereas Hamiltonian (6)
is non-Hermitian; the corresponding $S$-matrix $S=1+iT(U_{\bar{n}})$ is essentially
non-unitary. As a result, in the above-mentioned model the effect of antineutron
absorption acts in the opposite (wrong) direction [1,2]. The importance of unitarity
condition is well known [3-5]. Nevertheless, the model based on (5) and (6) is
frequently used (see below) because it greatly simplifies the calculation.

In this connection we continue consideration of some aspects of unitary models of
multistep processes in the nuclear matter. The decay channels (1) and (2) were
selected on the following reason. Let us consider the $n\bar{n}$ transition [6-8]
in the medium followed by annihilation
\begin{equation}
n\rightarrow \bar{n}\rightarrow M.
\end{equation}
This is a simplest process involving intermediate-state interaction since the
$n\bar{n}$ transition vertex corresponds to 2-tail diagram (see Fig. 1b). As
a result, Eqs. (5)-(7) lead to the simple formula $\Gamma _t=-2Im\epsilon
G_m\epsilon $ ($\epsilon $ is off-diagonal mass [7,8]), as opposed to the decay
$a\rightarrow \pi ^0\bar{n}$ shown in Fig. 1a. Besides, this process is of
independent interest [9]. (We note that all standard calculations of process (8)
are based on Eqs. (5) and (6) (see [8,9] for future references), which is wrong on
the reason given above.) Decays (1) and (2) are considered because they involve
the same $\bar{n}$-medium interaction as process (8).

For the decay $a\rightarrow \pi ^0\bar{n}$ in the medium we calculate the
$\Gamma _a$, $\Gamma _t$ and the branching ratio of channels using the 
low-density approximation [10] for the $\bar{n}$-medium interaction.
The results obtained are generalized to the decay
\begin{equation}
a\rightarrow b+c
\end{equation}
in the medium and reaction on a nucleon of the medium
\begin{equation}
a_1+N\rightarrow b+c
\end{equation}
followed by elastic and inelastic $c$-medium interactions.

We calculate the width of the decay (1). Our plan is as follows. 1) First of all
we construct and study the simplest process model suitable to the concrete calculations.
2) In the framework of this model we calculate the values listed above.

In order to construct the model correctly, we consider first the free-space $\bar{n}N$
annihilation (see Fig. 2a) and the process on a free nucleon
\begin{equation}
a+N\rightarrow \pi ^0+\bar{n}+N\rightarrow \pi ^0+M,
\end{equation}
shown in Fig. 2b (the free-space subprocess). The amplitude of free-space $\bar{n}N$
annihilation $M_a$ is defined as
\begin{equation}
<\!M\!\mid T\exp (-i\int dx{\cal H}_{\bar{n}N}(x))-1\mid\! \bar{n}N\!>=
N_a(2\pi )^4\delta ^4(p_f-p_i)M_a.
\end{equation}
Here ${\cal H}_{\bar{n}N}$ is the Hamiltonian of the $\bar{n}N$ interaction,
$N_a$ includes the normalization factors of the wave functions.
$M_a$ involves all the $\bar{n}N$ interactions followed by annihilation
including the {\em $\bar{n}N$ rescattering in the initial state}.

\begin{figure}[h]
  {\includegraphics[height=.25\textheight]{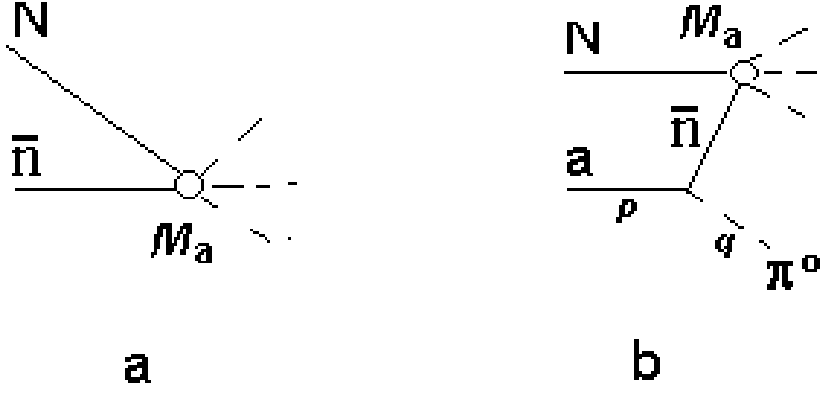}}
  \caption{{\bf a} Free-space $\bar{n}N$ annihilation. {\bf b} Free-space
reaction $a+N\rightarrow \pi ^0+\bar{n}+N\rightarrow \pi ^0+M$.}
\end{figure}

We write the formulas corresponding to Fig. 2b. The interaction Hamiltonian is
\begin{equation}
{\cal H}_I={\cal H}_d+{\cal H}_{\bar{n}N}.
\end{equation}
In the lowest order in ${\cal H}_d$ the process amplitude is given by
\begin{equation}
M_{2b}=gGM_a,
\end{equation}
\begin{equation}
G=\frac{1}{(p_0-q_0-m)-({\bf p}-{\bf q})^2/2m+i0},
\end{equation}
where $p$ and $q$ are the 4-momenta of the $a$-particle and $\pi ^0$, respectively,
$M_a$ is given by (12). Since $M_a$ contains all the $\bar{n}N$ interactions
followed by annihilation, the antineutron propagator $G$ is bare.

Consider now the decay (1) (see Fig. 3a). The background $a$-particle potential
is included in the wave function of $a$-particle $\Psi _a(x)$. The interaction
Hamiltonian is given by (3). In the lowest order in ${\cal H}_d$ the process
amplitude $M_{3a}$ is uniquely determined by the Hamiltonian (3):
\begin{equation}
M_{3a}=gGM_a^m.
\end{equation}
The amplitude of the $\bar{n}$-medium annihilation $M_a^m$ is given by
\begin{equation}
<\!f\!\mid T\exp (-i\int dx{\cal H}(x))-1\mid\! 0\bar{n}_{p-q}\!>=
N(2\pi )^4\delta ^4(p_f-p_i)M_a^m
\end{equation}
(compare with (12)). Here $\mid\! 0\bar{n}_{p-q}\!>$ and $<\!f\!\mid $ are the
states of the medium containing the $\bar{n}$ with the 4-momentum $p-q$ and
annihilation products, respectively; $N$ includes the normalization factors of
the wave functions.

\begin{figure}[h]
  {\includegraphics[height=.25\textheight]{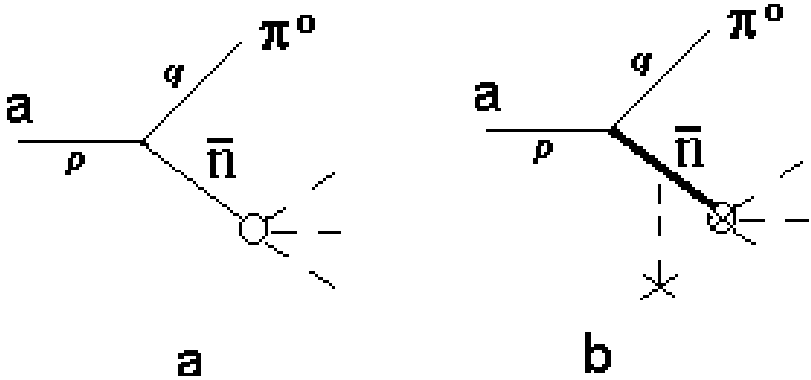}}
  \caption{{\bf a} Decay $a\rightarrow \pi ^0+\bar{n}\rightarrow \pi ^0+M$
in nuclear matter. The antineutron annihilation is shown by a circle. {\bf b}
The same as {\bf a}, but the antineutron propagator is dressed (see text).}
\end{figure}

The definition of the annihilation amplitude $M_a^m$ through Eqs. (17) is
natural. If the number of particles of medium is equal to unity, Eq. (17)
goes into (12). The antineutron annihilation width $\Gamma $ is expressed
through $M_a^m$ (see Eq. (24)). $M_a^m$ involves all the $\bar{n}$-medium
interactions followed by annihilation {\em including} the antineutron
rescattering in the initial state. Due to this, the propagator $G$ is bare.
As in (14), the antineutron self-energy $\Sigma =0$ since the interaction,
which can generate $\Sigma $, is involved in $M_a^m$.

If the Hamiltonian ${\cal H}$ is expressed through the $\bar{n}N$- and
$NN$-interactions, the amplitude $M_a^m$ contains in medium $\bar{n}N$-amplitudes
and dressed propagators. In this case the following condition should be fulfilled:
if $\rho \rightarrow 0$, the propagator is not dressed. However, our purpose is to
study the general features of the simplest model {\em suitable to the concrete
calculations} and not the medium effects. On this reason we consider the model (16)
which contains the block $M_a^m$ corresponding to the observable values. We would
like to emphasize this paragraph.

The fact that the propagator is bare is important and so we study this point in detail.
Using the same Hamiltonian (3), we try to construct the model with the dressed propagator
(see Fig. 3b). Denote
\begin{equation}
V={\rm Re}U_{\bar{n}}.
\end{equation}
As in (6), the $U_{\bar{n}}$ is the optical potential of $\bar{n}$. We recall the 
Hamiltonian ${\cal H}$ involves all the $\bar{n}$-medium interactions. 
In the Hamiltonian ${\cal H}$ we separate out the real potential $V$:
\begin{equation}
{\cal H}=V\Psi ^+_{\bar{n}}\Psi _{\bar{n}}+{\cal H}'
\end{equation}
and include it in the antineutron Green function
\begin{equation}
G_d=G+GVG+...=\frac{1}{G^{-1}-V}.
\end{equation}
The amplitude $M_{3b}$ of the process shown in Fig. 3b is
\begin{equation}
M_{3b}=gG_dM'_a.
\end{equation}
Since the amplitudes $M_{3a}$ and $M_{3b}$ correspond to one and the
same Hamiltonian (3), $M_{3a}=M_{3b}$ and
\begin{equation}
G_dM'_a =GM_a^m.
\end{equation}
The propagator $G_d$ is dressed: $\Sigma =V\neq 0$. According to (22), the
expressions for the propagator and vertex function are uniquely connected
(if $H_I$ is fixed). The "amplitude" $M'_a(V,H')$ should describe the
annihilation. However, below is shown $M'_a$ and model (21) are unphysical.
Comparing the left- and right-hand sides of (22), we see the following.

(1) If the number of particles of medium $n$ is equal to unity, model (21)
does not describe the free-space process shown in Fig. 2b because Eq. (14)
contains the bare propagator.

(2) The observable values ($\Gamma $, for example) are expressed through $M_a^m$
and not $M'_a$. Compared to $M_a^m$, $M'_a$ is truncated because the portion of
the Hamiltonian $H$ is included in the $G_d$. $M'_a$ has not a physical meaning.

(3) Amplitude (21) cannot be naturally obtained from the formal expansion of the
$T$-operator $T\exp (-i\int dx(V\bar{\Psi }_{\bar{n}}\Psi _{\bar{n}}+{\cal H}'))$.

(The formal expression for the dressed propagator should contain the annihilation
loops as well. In this case the statements given in pp. (1) and (2) are only
enhanced. A particle self-energy should be considered in the {\em context of
the concrete problem}. The dressed propagator arises naturally if $V$ and
${\cal H}'$ are the principally different interactions and vertex function
does not depend on $V$. In our problem one and the same field generates
$\Sigma $ and $M_a^m$.)

(4) Equations (20) and (21) mean that the annihilation is turned on upon forming
of the self-energy part $\Sigma =V$ (after multiple rescattering of $\bar{n}$).
This is counter-intuitive since at the low energies [11,12]
\begin{equation}
\frac{\sigma _a}{\sigma _t}>0.7
\end{equation}
($\sigma _t=\sigma _a+\sigma _s$, $\sigma _a$ and $\sigma _s$ are the cross
sections of free-space $\bar{n}N$ annihilation and scattering, respectively)
and inverse picture takes place: in the first stage of $\bar{n}$-medium
interaction the annihilation occurs.

The realistic {\em competition} between scattering and annihilation should
be taken into account. Both scattering and annihilation vertices should
occur on equal terms in $M_a^m$ or $G_d$. According to pp. (1)-(3) the
latest possibility should be excluded. Model (16) is free from
drawbacks given in pp. (1)-(3). It reproduces the ratio (23) as well.

\begin{figure}[h]
  {\includegraphics[height=.25\textheight]{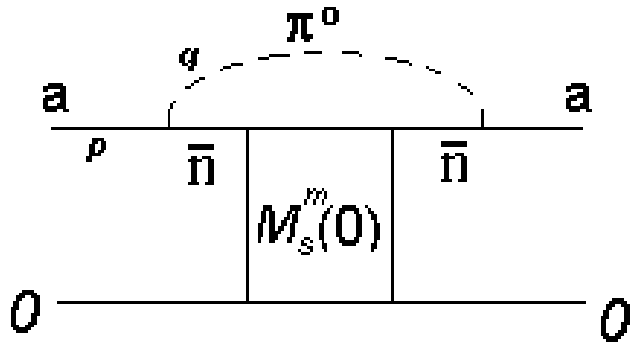}}
  \caption{On-diagonal matrix elements corresponding to the process
$(a-\mbox{medium})\rightarrow \pi ^0+(\bar{n}-\mbox{medium})\rightarrow
(a-\mbox{medium})$.}
\end{figure}

Indeed, we calculate the width of the decay (1) shown in Fig. 3a. Put ${\bf p}=0$
for simplicity. The pion wave function is $\Phi (x)=(2q_0\Omega )^{-1/2}\exp (-iqx)$.
Using the amplitude (16), the decay width $\Gamma _a$ is found to be
\begin{equation}
\Gamma _a=\int d{\bf q}\frac{1}{2q_0(2\pi )^3}g^2G^2\Gamma ,
\end{equation}
where $\Gamma $ is the annihilation rate of the $\bar{n}$ with the 4-momentum $p-q$.
The corresponding amplitude $M_a^m$ is given by (17).

The number of exchanges with the medium or particle, which appears in (20), is very
important but unobservable value. In particular, it is responsible for the particle
self-energy and process suppression by the potential $V$. The low-density limit enables
one to verify directly the condition (23). Indeed, in the low-density approximation
$\Gamma =v\rho \sigma _a$ and
\begin{equation}
\Gamma _a=\int d{\bf q}F_1\sigma _a,
\end{equation}
\begin{equation}
F_1=\frac{g^2G^2v\rho }{2q_0(2\pi )^3},
\end{equation}
$q_0^2={\bf q}^2+m_{\pi }^2$, where $m_{\pi }$ is the pion mass.

Let us calculate the total width $\Gamma _t$ of the decay $a\rightarrow \pi ^0\bar{n}$
in the medium. In the lowest order in ${\cal H}_d$ the on-diagonal matrix
element shown in Fig. 4 is given by
\begin{equation}
M_{3}=\int \frac{dq}{(2\pi )^4}g\frac{i}{q^2-m_{\pi }^2+i0}GM_s^m(0)Gg.
\end{equation}
Here $M_s^m(0)$ is the forward scattering amplitude of $\bar{n}$ in the medium.
We integrate over $q_0$ and use the optical theorem in the left- and right-hand
sides of (27):
\begin{equation}
\frac{1}{T_0}2{\rm Im}M_{3}=\Gamma _t,
\end{equation}
\begin{equation}
\frac{1}{T_0}2{\rm Im}M^m_s(0)=v\rho \sigma _t
\end{equation}
($T_0$ is the normalization time, $T_0\rightarrow \infty  $), resulting in
\begin{equation}
\Gamma _t=\int d{\bf q}F_1\sigma _t.
\end{equation}
Denoting $\mid\!{\bf q}\!\mid =q$, one gets finally
\begin{equation}
r=\frac{\Gamma _a}{\Gamma _t}=\frac{\int dqF(q)\sigma _a(q)}{\int dqF(q)\sigma
_t(q)},
\end{equation}
\begin{equation}
F=\frac{q^3G^2}{q_0}.
\end{equation}
In the differential form
\begin{equation}
\frac{d\Gamma _a/dq}{d\Gamma _t/dq}=\frac{\sigma _a}{\sigma _t}>0.7,
\end{equation}
where Eq. (23) was used. In a similar manner, one obtains
\begin{equation}
\frac{d\Gamma _s/dq}{d\Gamma _t/dq}=\frac{\sigma _s}{\sigma _t}<0.3.
\end{equation}
For the description of the intermediate-state $\bar{n}$-medium interaction
in (25), (29) and (34) the low-density approximation has been used.

We consider the free-space decay (9). Let for the decay in the medium the
inelastic
\begin{equation}
a\rightarrow b+c\rightarrow b+x
\end{equation}
and elastic
\begin{equation}
a\rightarrow b+c\rightarrow b+c
\end{equation}
interactions of the $c$-particle take place. Here $c\rightarrow x$ and $c\rightarrow
c$ imply the reaction induced by $c$-particle and scattering of $c$-particle in the
medium, respectively. Let $\Gamma _r$ and $\Gamma _s$ be the widths of decays (35)
and (36), respectively; $\sigma _r$ and $\sigma _s$ are the cross sections of
free-space $cN$ reaction and scattering, respectively; $\Gamma _t=\Gamma _r+\Gamma _s$
and $\sigma _t=\sigma _r+\sigma _s$; $q$ is the 4-momenta of the $b$-particle. With the 
replacement $\Gamma _a\rightarrow \Gamma _r$ and $\sigma _a\rightarrow \sigma _r$, the 
relations (31)-(34) describe the decay channels (35) and (36).

For the reaction (10) in the medium instead of (35) and (36) we consider the channels:
\begin{equation}
a_1+N\rightarrow b+c\rightarrow b+x
\end{equation}
and
\begin{equation}
a_1+N\rightarrow b+c\rightarrow b+c.
\end{equation}
In this case the similar relations take place as well. The corresponding changes in
$F$ are minimum and non-principal. For example, for the channels (37) and (38) we get
\begin{equation}
\frac{d\sigma _r^m/dq}{d\sigma _s^m/dq}=\frac{\sigma _r}{\sigma _s},
\end{equation}
where $\sigma _r^m$ and $\sigma _s^m$ are the cross sections of the reactions
(37) and (38) in the medium, respectively.

Relations (33), (34) relate the in medium values $d\Gamma _a/dq$, $d\Gamma _s/dq$
and $d\Gamma _t/dq$ with the free-space branching ratio of channels
$\sigma _a/\sigma _t$ and $\sigma _s/\sigma _t$. The similar relations (Eq. (39),
for example) take place for the decays (35), (36) and reactions (37), (38) in the
medium. The nuclear medium changes the amplitudes [13] and branching ratio of
channels [14] in hadron-nucleon interactions. Besides, the strong antineutron
absorption takes place. Due to this, at the nuclear densities $\Gamma _a/\Gamma _t
\approx 1$. However, if $\rho \rightarrow 0$, the relations (31)-(34) and (39)
should be reproduced.

The definition of annihilation amplitude through Eqs. (12) and (17) is natural
since it corresponds to the observable values. In that event the intermediate
particle propagator is bare. In this sense the above-considered model is opposite
to that shown in Fig. 1 which contains the full in medium propagator calculated
through the Hermitian Hamiltonian. In the intermediate cases one can get a double
counting or opposite error. As a rule, an additional self-energy leads to
suppress the process and so we would like to call particular attention to this
point as well as the unitarity (see paragraph below Eq. (7)).

Finally, once the amplitudes are defined by (12) and (17), the propagator is bare;
the processes on a free nucleon shown in Fig. 2 are reproduced; for the branching
ratio of channels the relations (33), (34) and (39) take place. In fact, these
relations should be fulfilled for any process model and can be considered as
necessary condition for the correct model construction.

\end{document}